\documentclass[12pt]{article}
\usepackage[utf8]{inputenc}
\usepackage{textcomp}
\usepackage{graphicx}
\usepackage{listings}
\usepackage{xcolor}
\usepackage{hyperref}
\title{Client-Side Zero-Shot LLM Inference for Comprehensive In-Browser URL Analysis}
\author{Avihay Cohen}
\date{Jun 2025}

\begin{document}

\maketitle

\begin{abstract}
Malicious websites and phishing URLs pose an ever-increasing cybersecurity risk, with phishing attacks growing by 40\% in a single year~\cite{Kaspersky2024}. Traditional detection approaches rely on machine learning classifiers or rule-based scanners operating in the cloud, but these face significant challenges in generalization, privacy, and evasion by sophisticated threats. In this paper, we propose a novel \emph{client-side} framework \cite{url_analysis} for comprehensive URL analysis that leverages zero-shot inference by a local large language model (LLM) running entirely in-browser. Our system uses a compact LLM (e.g., 3B/8B parameters) via WebLLM\cite{Ruan2024WebLLM} to perform reasoning over rich context collected from the target webpage, including static code analysis (JavaScript abstract syntax trees, structure, and code patterns), dynamic sandbox execution results (DOM changes, API calls, and network requests), and visible content. We detail the architecture and methodology of the system, which combines a real browser sandbox (using iframes) resistant to common anti-analysis techniques, with an LLM-based analyzer that assesses potential vulnerabilities and malicious behaviors without any task-specific training (\emph{zero-shot}). The LLM aggregates evidence from multiple sources (code, execution trace, page content) to classify the URL as benign or malicious and to provide an explanation of the threats or security issues identified. We evaluate our approach on a diverse set of benign and malicious URLs, demonstrating that even a compact client-side model can achieve high detection accuracy and insightful explanations comparable to cloud-based solutions, while operating privately on end-user devices. The results show that client-side LLM inference is a feasible and effective solution to web threat analysis, eliminating the need to send potentially sensitive data to cloud services. We discuss how our approach mitigates evasion techniques, the trade-offs in performance when using smaller models, and why scaling analysis via on-device models is preferable to cloud-scale LLMs for this domain.
\end{abstract}

\tableofcontents

\section{Introduction}
The web browser is the primary interface for users interacting with online content, making it a common vehicle for cyber attacks such as phishing, drive-by downloads, and malicious scripts. Recent reports indicate a sharp rise in phishing and malicious URL campaigns - for example, phishing attacks increased by 40\% in 2023 alone~\cite{Kaspersky2024}. This escalation in threats places urgency on effective URL analysis and web content scanning techniques. Although numerous machine learning and deep learning based models have been proposed for malicious URL detection~\cite{Jain2019,Le2018URLNet}, these approaches often suffer from generalization issues and can be outmaneuvered by determined adversaries. Attackers frequently employ client-side cloaking and dynamic content generation to show benign content to security scanners while delivering malicious payloads to real users~\cite{Zhang2021}. This gap between the scanning environment and a real user’s browsing environment allows many threats to slip past traditional detectors.

A promising new direction is to leverage the reasoning ability of large language models (LLMs) to perform security analysis. LLMs have demonstrated an ability to reason in a zero-shot manner, solving tasks without task-specific training when given proper prompts~\cite{Kojima2022}. Recent studies suggest that LLMs can act as one-shot or few-shot URL classifiers and even provide explanations for their decisions by synthesizing knowledge of known attack patterns with observed data. However, deploying LLMs for web content analysis at scale introduces significant challenges. Cloud-based LLM services raise privacy concerns, since the content of a suspect webpage (which may include private user data or sensitive information) must be sent to a third-party server for analysis. Furthermore, relying on cloud LLM APIs for every URL visit is costly and does not scale well when millions of URLs or user browsing sessions need to be evaluated in real-time. Adversaries may also attempt to evade cloud-based analysis by fingerprinting known scanner infrastructure or IP ranges, a technique observed in advanced phishing kits~\cite{Zhang2021}.

In this work, we address these challenges by bringing the power of LLM inference directly into the client's browser. We introduce a system for \textbf{comprehensive in-browser URL analysis using zero-shot LLM inference on the client side}\cite{url_analysis}. Our approach is comprehensive in that it analyzes a suspect URL from multiple angles:
\begin{itemize}
\item \textit{Static analysis of JavaScript code}: The system fetches all scripts from the webpage and parses them into abstract syntax trees (ASTs) to extract structured metadata such as function names, variables, loops, strings, and patterns of API usage. This static analysis component flags suspicious code constructs (e.g., use of \texttt{eval} on obfuscated strings, event handlers for keystroke logging, suspicious URL patterns in network calls) without executing the code.
\item \textit{Dynamic analysis via sandboxed execution}: The URL is loaded in a real browser environment using an advanced sandbox framework that leverage isolated iframes. The system intercepts and monitors dynamic behavior, including changes to the Document Object Model (DOM), calls to runtime Web APIs (such as network fetches, redirects, cookie access, and geolocation requests), and any content that becomes visible to the user. By executing the page in an environment that closely mirrors a user's browser (including using a typical user-agent string and screen dimensions), we capture the runtime behavior and UI rendered, defeating techniques that rely on detecting headless bots or non-interactive crawlers. Our sandbox leverages standard web technologies (iframes and Proxy objects) with careful instrumentation to remain resistant to common evasion tactics.
\item \textit{LLM-based reasoning and threat assessment}: A local LLM running entirely in the browser (via WebGPU and WebAssembly, using the WebLLM framework~\cite{Ruan2024}) takes the collected evidence from static and dynamic analysis as input. We craft multiple prompts that includes the URL’s key characteristics (domain, path, etc.), the extracted code features (e.g., list of function names, presence of suspicious patterns), summary of dynamic findings (e.g., “the page created an invisible iframe and made network requests to \texttt{evil.com} API”), and the visible text/content of the page. Without any fine-tuning, the LLM is asked in a zero-shot manner to assess whether the URL/page is malicious or benign, explain the reasoning, and identify any specific vulnerabilities or malicious behaviors present. Because the model is running locally, this analysis is privacy-preserving (no data leaves the user's machine) and nearly real-time.
\end{itemize}

By combining these components, our system performs a holistic analysis akin to what a human security expert might do: inspect the page’s code, observe its behavior, and read its content, then make an informed judgment about its intent. All of this occurs within the user's browser client, on the fly as a URL is analyzed, eliminating the need for cloud queries.

We implement our approach with a focus on using relatively compact LLMs that are feasible to deploy in a browser. In particular, we experiment with models in the 2B--8B parameter range (such as a LLaMA-family 8B model, a 3B distilled model, and recent efficient models like Microsoft’s \texttt{Phi-3} and Google’s \texttt{Gemma-2B} small LMs). These models, especially when quantized (e.g., 4-bit weights), can be loaded and run with hardware acceleration in modern browsers via WebGPU, achieving reasonable throughput~\cite{Ruan2024}. Despite their smaller size compared to state-of-the-art 100B+ parameter models, recent work has shown that these small models can remain surprisingly capable on specialized tasks~\cite{IBM2024Gemma}. They also run faster and can be used without network connectivity or ongoing API costs.

The contributions of this paper are as follows:
\begin{itemize}
\item We design a novel \textbf{client-side URL analysis architecture} that integrates static code parsing, dynamic sandbox execution, and LLM-based reasoning entirely within the browser. To our knowledge, this is the first framework to perform comprehensive web threat analysis using an in-browser LLM inference engine.
\item We develop techniques for \textbf{JavaScript AST analysis and behavior tracing} 
in untrusted webpages, extracting features that feed into the LLM. We describe how our system hooks and monitors web API calls and DOM updates in a sandboxed iframe to detect stealthy behaviors, while minimizing the footprint that might reveal the presence of an analysis tool to the page.
\item We demonstrate the feasibility of \textbf{zero-shot LLM inference for security analysis} using a compact model running in-browser. Our methodology requires no task-specific training or labeled data for the LLM; it relies on the model’s pre-trained knowledge of malicious patterns and its reasoning ability to correlate static and dynamic observations.
\item We evaluate the system on a dataset of benign and malicious URLs, measuring detection performance and highlighting examples of the LLM’s explanatory outputs. We compare different model sizes (8B vs 3B vs 2B) and architectures, analyzing the trade-offs in detection accuracy and runtime performance on the client side. We also compare against a traditional machine-learning URL classifier baseline to illustrate the advantages of our approach in terms of explanation and resilience to evasion.
\item We provide a thorough discussion of \textbf{why client-side inference is advantageous} in this domain. We analyze the limitations of cloud-based analysis (privacy, scalability, evadability) and show how our approach addresses these. We discuss current limitations of our system (such as the performance gap of smaller models and potential evasion if attackers specifically target LLM weaknesses) and outline future work to further improve client-side threat detection.
\end{itemize}

The remainder of the paper is organized as follows. Section~\ref{sec-related} reviews related work in malicious URL detection, web malware analysis, and the emerging use of LLMs in security. Section~\ref{sec-architecture} describes the architecture of our client-side analysis system. Section~\ref{sec-methodology} details the methodology, including the static analysis pipeline, dynamic sandbox instrumentation, and the LLM prompt engineering. In Section~\ref{sec-experiments}, we present our experimental setup, datasets, and baseline comparisons, and in Section~\ref{sec-results} we report the evaluation results. Section~\ref{sec-discussion} provides a discussion on the implications of our findings, including the feasibility of client-side deployment and the comparison to cloud-based approaches. Finally, Section~\ref{sec-conclusion} concludes the paper.

\section{Related Work}\label{sec-related}
\subsection{Malicious URL and Web Content Detection}
Malicious URL detection has been a long-standing research area in cybersecurity. Early approaches focused on lexical features of URLs and blacklisting, while later methods employed machine learning on features extracted from URLs, HTML content, and network characteristics. For example, Jain and Gupta~\cite{Jain2019} proposed a machine learning approach for phishing detection using features from hyperlinks (URL tokens, domain reputation, etc.). With the rise of deep learning, researchers developed models that directly learn URL representations, such as URLNet by Le~\textit{et al.}~\cite{Le2018URLNet}, which uses convolutional neural networks to embed URLs for classification. These systems can achieve good accuracy on known datasets, but they often struggle with generalization to new phishing campaigns or novel attack techniques. One common issue is that static classifiers can be evaded by adversaries who change superficial features of URLs or page content (e.g., typosquatting new domains or injecting random noise strings to confuse classifiers).

Another line of work has examined the content of webpages for malicious indicators. Traditional anti-phishing toolbars and browser filters would look for known bad keywords or images. More advanced systems like Cova \textit{et al.}'s approach~\cite{Cova2010} combined static analysis with dynamic emulation to detect drive-by download attacks in pages. They ran JavaScript code in an instrumented environment to catch malicious behaviors like heap spraying. Similarly, \textit{Zozzle} by Curtsinger \textit{et al.}~\cite{Curtsinger2011} introduced a mostly static JavaScript malware detector operating within the browser, using a statistical model over AST-derived features to identify malware code. Zozzle demonstrated that fast in-browser scanning of JavaScript is possible with low overhead, though its pattern-matching approach could be circumvented by heavily obfuscated scripts.

Our work builds on insights from these prior works: purely static or lexical analysis is insufficient against adaptive threats, and in-browser scanning is feasible and can be improved with modern techniques. We extend the concept of in-browser analysis by incorporating an LLM to reason over the combined static and dynamic evidence, something that was not available to earlier systems.

\subsection{Evasion and Dynamic Analysis Techniques}
Attackers have developed sophisticated evasion techniques to defeat automated analysis. One prominent method is \textbf{client-side cloaking}, where malicious pages detect the absence of typical user interactions or the presence of known crawler environment clues and then alter their behavior~\cite{Zhang2021}. For instance, a phishing page might load a benign decoy page if it suspects it is being scanned by an anti-phishing service (which might be indicated by no mouse movements, an unusual User-Agent string, or known headless browser signatures). Zhou \textit{et al.}'s \emph{CrawlPhish} study~\cite{Zhang2021} extensively documented how phishing kits employ client-side checks (like requiring a real click or keyboard input, or checking if the browser has a graphical canvas to detect headless Chrome) to cloak their malicious payloads from crawlers. Their findings underscore the need for using an analysis environment that is indistinguishable from a real user on a real browser.

Dynamic analysis sandboxes, such as those used by services like Google Safe Browsing and various malware analysis sandboxes (e.g., Any.Run, JoeSandbox), attempt to load and run webpages in virtual browsers to observe malicious behavior. While effective to an extent, these cloud-based sandboxes can be resource-intensive and still prone to detection. Moreover, many such services operate after the fact (scanning URLs submitted to them) rather than proactively scanning every page a user visits in real time, due to scalability limits.

Our approach adopts dynamic analysis in a \emph{distributed client-side} fashion. Instead of one cloud service analyzing thousands of URLs (which can be detected and rate-limited by attackers, and is costly), each user's browser can perform dynamic analysis on the few URLs that user is about to visit, using the user's own computing resources. By leveraging a real browser iframe within the user's session, we naturally reproduce a genuine browsing context (with real user-agent, IP, locale, etc.), making it much harder for the malicious page to identify the analysis. We also incorporate techniques like triggering minimal user-like interactions if needed (e.g., programmatically dispatching click events or simulating time delays) to satisfy simple cloaking checks.

An additional benefit of client-side analysis is privacy: prior dynamic analysis frameworks required uploading the URL (and potentially page content) to a third-party service. In scenarios where the URL could contain private tokens or the page requires login, cloud analysis is not acceptable. Our system keeps the entire process local to the user.

\subsection{Large Language Models for Security Analysis}
The use of LLMs in security is an emerging area of research. Large language models like GPT-3/GPT-4 have shown impressive ability to understand code and even detect vulnerabilities or malicious intent when carefully prompted. For example, OpenAI's GPT-4 Technical Report noted the model can assist in code auditing tasks with high competency. There have been works exploring LLMs as explainers for predictions (e.g., Kroeger \textit{et al.} studied whether LLMs can serve as post-hoc explainers of ML decisions). Closer to our scope, some recent studies have treated LLMs themselves as classifiers for malicious content, taking advantage of their broad knowledge. These studies often show that even without training on specific security datasets, an LLM can make reasonable judgments about, say, whether a given piece of code is suspicious or a given URL looks phishy, especially when the prompt provides contextual information and instructions.

Our approach is aligned with this trend, using an LLM as the core analysis engine. However, a key distinction is we run the LLM locally within the browser environment. The WebLLM project~\cite{Ruan2024} demonstrated that it is possible to achieve near-native speeds for LLM inference in the browser via WebGPU, and our work leverages that capability. We also tailor the prompt engineering to the threat detection task, effectively building a zero-shot classifier with explanation.

In summary, our system integrates techniques from static code analysis within the browser, a novel dynamic sandbox emulator, and LLM-based reasoning to create a novel solution for client-side URL analysis. To the best of our knowledge, this is the first work to integrate all these components and evaluate the feasibility of running it entirely in a user's browser at runtime.

\section{Architecture}\label{sec-architecture}
\begin{figure}[h!]
\centering
\includegraphics[
  width=\textwidth,
  height=0.6\textheight,
  keepaspectratio
]{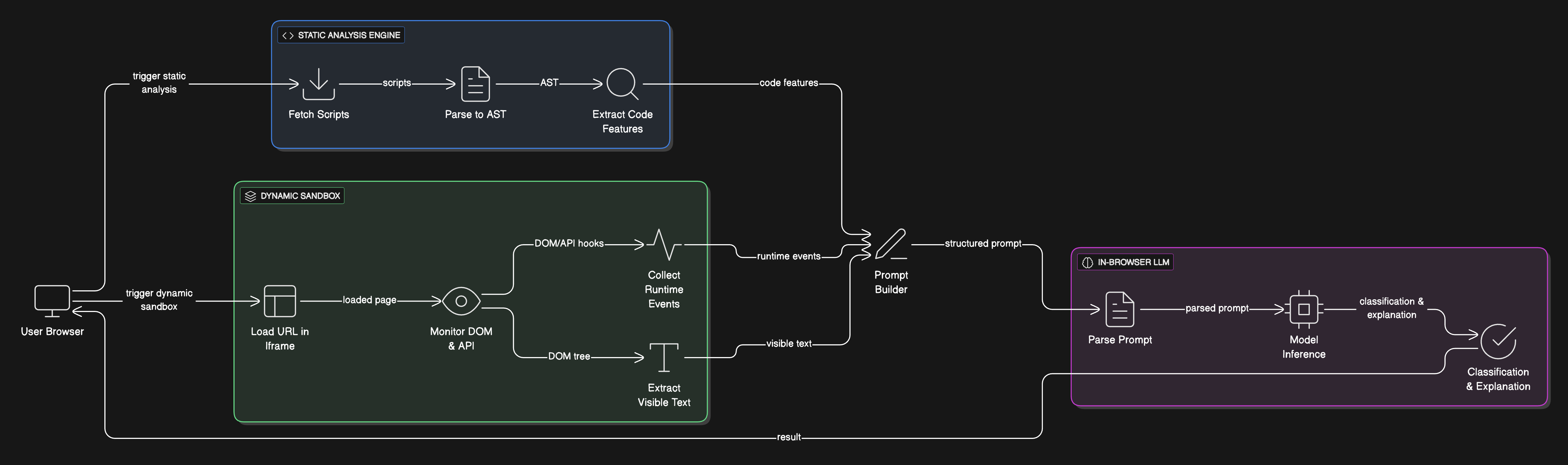}
\fbox{\parbox{0.95\textwidth}{\centering \textbf{ Figure~1: System Architecture.}
A high-level diagram showing the flow: The user's browser triggers analysis for a URL. The Static Analysis Engine fetches and parses scripts to AST, extracting code features. The Dynamic Sandbox loads the URL in an iframe, with hooks monitoring DOM changes and API calls. Collected data (code features, runtime events, visible text) are sent to the Prompt Builder, which constructs a detailed prompt. The in-browser LLM (WebLLM) processes this prompt and outputs a classification (malicious/benign) with explanation.}}
\label{fig-architecture}
\end{figure}

Our client-side URL analysis system is composed of several coordinated components that together perform end-to-end threat detection and explanation. Figure~\ref{fig-architecture} provides an overview of the architecture. The major components include: (i) the \textbf{Static Analysis Engine} for JavaScript parsing and feature extraction, (ii) the \textbf{Dynamic Execution Sandbox} for loading the page in an instrumented environment, (iii) the \textbf{Data Aggregation and Prompt Builder} which compiles the analysis results into multiple prompts, and (iv) the \textbf{LLM Inference Module} (running via WebLLM in-browser) which processes the prompts and generates an analysis response. A frontend UI is  provided to the user (e.g., a browser extension popup) to review the results, but the core analysis runs automatically when triggered for a given URL.

The system operates as follows: when a user (or an automated background process) initiates analysis of a target URL, a new isolated iframe (or a new browser context) is created to load the page. The Static Analysis Engine, in parallel, downloads the JavaScript files of the page (either by intercepting network responses or via the browser's developer API if available or through a cors proxy) and performs a static scan. The engine is implemented in TypeScript, using Acorn to parse scripts into ASTs. It then traverses the AST to collect various metadata:
\begin{itemize}
\item \textbf{Identifiers:} List of function names, variable names, and literal constants present in the code. These can reveal suspicious naming patterns (e.g., a function named `hookKeyboard` or a variable `bankPassword` is suspicious in a random page).
\item \textbf{Structural features:} Presence of constructs like \texttt{eval()}, \texttt{Function()} constructor, dynamic script injection 
\texttt{document.createElement('script')}, event listeners (onmousedown, onkeypress), loops that iterate over large arrays (could indicate decoding loops), etc.
\item \textbf{String patterns:} Any hardcoded URL substrings, suspicious domain names, long hexadecimal or Base64 strings (often used in obfuscated payloads), or known malware signatures if any.
\item \textbf{Control flow indicators:} The AST can be analyzed for deep nesting or dead code (some malware uses opaque predicates), and for whether the code is largely minified/obfuscated (e.g., many single-letter variable names or lack of whitespace can be a sign).
\item \textbf{Obfuscation Detection} - The AST is analyzed for common obfuscation techniques.

\end{itemize}
The Static Analysis Engine produces a structured summary of each script. For example, it may output a JSON-like structure for each file: listing its functions and variables, and flagging any detected patterns (e.g., "uses eval on string literal", "calls \texttt{atob()} decoding", "creates hidden form element","API keys detected in code","private/public encryption keys").

Meanwhile, the Dynamic Execution Sandbox is responsible for observing the page’s runtime behavior. We leverage a sandboxed iframe that runs in the context of an emulated origin. To instrument it, our system injects a small script into the iframe as it loads (for instance, by setting the iframe's src to a data URL that bootstraps a loader script, which then render the target html while retaining full execution control, or by using the browser extension content script capabilities to pre-inject hooks). The instrumentation script hooks into key browser APIs by monkey-patching them. For example, we override \texttt{window.fetch}, \texttt{XMLHttpRequest.prototype.send}, \texttt{WebSocket}, and other networking functions to log any outbound requests (recording the URL, method, and perhaps payload size, but not blocking them). We similarly hook \texttt{document.cookie} accessors, \texttt{localStorage} access, and other storage APIs to see if the script is trying to read/write sensitive data.

We also attach event listeners to capture DOM modifications. This can be done by observing mutations (using the \texttt{MutationObserver} API on the iframe's document body) and by hooking specific DOM methods like \texttt{appendChild} or \texttt{insertBefore}. We pay special attention to changes that make content visible or hidden. For example, if an element is created with CSS that places it off-screen or invisible, we note that (common in pop-up phishing forms or overlays). If a new login form or input field is added to the DOM, we capture that as well.

Finally, we capture the \textbf{visible text} on the page after a short delay (once network activity has settled). This is done by extracting all text nodes from the DOM that are not hidden by CSS. The visible text is important because it might contain clues like "Secure Login", brand names, or requests for credentials, which an LLM can recognize as phishing indicators when combined with other evidence.

The sandbox is designed to be \emph{resistant to detection}. We do not use any obvious headless mode (we rely on the user's actual browser, which passes all typical environment checks). The iframe can be made virtually invisible to the user (e.g., 0x0 pixels or hidden tab), but we ensure that scripts running in it do not detect unusual conditions. Our injected hooks try to be stealthy, for instance, replacing \texttt{window.fetch} with a wrapper that calls through to the original function, and making sure properties like \texttt{toString()} on the function return the expected native code string to avoid fingerprinting of the hook. We also handle timing: some malware checks how quickly certain actions happen (to detect automated environment). We can introduce slight random delays in our automated interactions and let the page mostly run at normal speed. In essence, the sandbox aims to let the page behave as if a user is visiting, while quietly recording its actions.

After a fixed analysis period (say 3-5 seconds, or earlier if the page finishes loading sooner and no new activity is observed), the system gathers all logged data from both static and dynamic analysis. This data is then passed to the Prompt Builder. The Prompt Builder composes a textual prompt for the LLM. The prompt typically has a predefined template, for example \emph{github.com} analysis:

\begin{small}
\begin{verbatim}
Q:

[System role: Security Analyst AI]
 Analyze sandbox API behavior for security risks.

API Calls Summary:
window.fetch: 1x, EventTarget.addEventListener: 100x,
newGlobalProperties: 76x, Document.querySelector: 44x, 
Document.getElementById: 25x, Document.createElement: 18x,
Document.querySelectorAll: 16x, Node.appendChild: 9x,
Node.cloneNode: 9x, EventTarget.removeEventListener: 8x, 
window.setTimeout: 6x

Risk Categories: low: 1 risks

Provide concise analysis as JSON:
{
  "sandboxRiskScore": <0-100>,
  "sandboxRiskLevel": "<Low|Medium|High|Critical>",
  "sandboxFindings": [{"title": "<issue>", "severity": "
  <Low|Medium|High|Critical>"}]
}

A:

 {
  "sandboxRiskScore": 20,
  "sandboxRiskLevel": "Low",
  "sandboxFindings": [
    {
      "title": "Potential Information Disclosure through
      newGlobalProperties",
      "severity": "Low"
    }
  ]
}

   \end{verbatim}
   \end{small}

\begin{small}
\begin{verbatim}
Q:

[System role: Security Analyst AI]
 Evaluate trust for: github.com

Indicators:
- SSL Certificate: Yes
- Code Quality Score: 99/100
- Domain: github.com

Provide trust assessment as JSON:
{
  "score": <0-100>,
  "level": "<Low|Medium|High>",
  "factors": ["<factor1>", "<factor2>"]
}

A:

  {
  "score": 92,
  "level": "High",
  "factors": ["SSL Certificate: Yes", "Code Quality Score: 99/100"]
}

   \end{verbatim}
   \end{small}

   \begin{small}
\begin{verbatim}
Q:

[System role: Security Analyst AI]
  Analyze JavaScript security for: wp-runtime-7ec44d86e5dd.js

Key indicators:
- Obfuscated: No
- Security issues: 2
- Invoked APIs: unknown.call, unknown.every, Object.keys, unknown.e,
o.splice,e.splice, a, s.d, Object.getPrototypeOf, Object.create, s.r,
_, e.indexOf, unknown.forEach,Object.getOwnPropertyNames, s.o,
Object.defineProperty,Promise.all, unknown.reduce, unknown.o,
Function,unknown.push, document.getElementsByTagName,
r.getAttribute, document.createElement, d.setAttribute, 
unknown.indexOf, clearTimeout, unknown.removeChild, a.forEach

- Defined Functions: s, m, e, _, o
- Dangerous APIs: 1

Provide analysis as JSON:
{
  "summary": "<brief security summary>",
  "securityAnalysis": {
    "riskLevel": "<Low|Medium|High|Critical>",
    "vulnerabilities": [
      {
        "type": "<vulnerability type>",
        "severity": "<Low|Medium|High|Critical>"
      }
    ]
  }
}

A:

 {
  "summary": "Low security risk found in wp-runtime-7ec44d86e5dd.js",
  "securityAnalysis": {
    "riskLevel": "Low",
    "vulnerabilities": [
      {
        "type": "Potential Insecure Deserialization",
        "severity": "Low"
      }
    ]
  }
}

   \end{verbatim}
   \end{small}

\begin{small}
\begin{verbatim}
Q:

[System role: Security Analyst AI]
 Analyze global properties created during sandbox execution for phishing 
 and malicious behavior:

URL: https://github.com
Domain: github.com
Total New Global Properties: 76

GLOBAL PROPERTIES DETECTED:
1. webpackChunk
2. IncludeFragmentElement
3. RemoteInputElement
4. ActionListElement
5. ActionBarElement
6. DetailsMenuElement
7. AnchoredPositionElement
8. DialogHelperElement
9. FocusGroupElement
10. ScrollableRegionElement
[truncated]....

CRITICAL INDICATORS TO ANALYZE:

1. DATA EXFILTRATION:
   - Properties containing "send", "post", "export", "transmit"
   - Functions for data collection and transmission
2. KEYLOGGING/MONITORING:
   - Properties for input monitoring
   - Keystroke capturing functions
   - User behavior tracking
3. SESSION HIJACKING:
   - Properties for token/cookie manipulation
   - Session management functions
   - Authentication bypass mechanisms
   
RISK ASSESSMENT CRITERIA:
- High Risk: Credential keywords + brand names + obfuscation
- Medium Risk: Suspicious patterns + function-like behavior
- Low Risk: Minimal suspicious patterns

Provide comprehensive global properties phishing analysis as JSON:
{
  "isPhishing": <true|false>,
  "confidence": <0-100>,
  "phishingType": "<none|data-exfiltrator|keylogger|session-hijacker>",
  "riskLevel": "<Low|Medium|High|Critical>",
  "globalPropIndicators": [
    {
      "property": "<property name>",
      "type": "<data-exfiltrator|keylogger|session-hijacker>",
      "description": "<why this property is suspicious>",
      "severity": "<Low|Medium|High|Critical>"
    }
  ],
  "behaviorAnalysis": {
    "hasDataExfiltrators": <true|false>,
    "hasKeyloggers": <true|false>,
    "hasSessionHijackers": <true|false>
  },
  "legitimacyScore": <0-100, lower means more suspicious>,
  "recommendation": "<detailed recommendation based on global properties analysis>",
  "globalPropsRiskScore": <0-100, higher means more likely phishing/malicious>
}
A:

{
  "isPhishing": false,
  "confidence": 80,
  "phishingType": "none",
  "riskLevel": "Low",
  "globalPropIndicators": [
    {
      "property": "webpackChunk",
      "type": "none",
      "description": "This property is a common library
      chunk and does not indicate phishing behavior.",
      "severity": "Low"
    },
    {
      "property": "IncludeFragmentElement",
      "type": "none",
      "description": "This property is a standard
      HTML element and does not indicate phishing behavior.",
      "severity": "Low"
    },
    {
      "property": "RemoteInputElement",
      "type": "none",
      "description": "This property is a standard
      HTML element and does not indicate phishing behavior.",
      "severity": "Low"
    },
    {
      "property": "DetailsMenuElement",
      "type": "none",
      "description": "This property is a standard
      HTML element and does not indicate phishing behavior.",
      "severity": "Low"
    }
  ],
  "behaviorAnalysis": {
    "hasDataExfiltrators": false,
    "hasKeyloggers": false
  },
  "legitimacyScore": 20,
  "recommendation": "github.com  is a well-known domain, and the
  provided properties do not
  contain anything abnormal or suspicious.",
  "globalPropsRiskScore": 30
}

   \end{verbatim}
   \end{small}

Those are real examples; the prompts are carefully formatted to be within the token limits of the model (which for a 8B model at 4k context might be 4096 tokens, sufficient for a summary of the page's content). We design the prompts to be as informative as possible while still relying on the LLM's own knowledge to connect the dots.

The \textbf{LLM Inference Module} then runs those prompts through the model. We use the WebLLM runtime~\cite{Ruan2024} to execute the model in a web worker with GPU acceleration. The model (e.g., a variant of LLaMA or CodeLlama, possibly fine-tuned on security/code domains) generates a response. The response typically includes a verdict (malicious or benign), an explanation highlighting which evidence led to that conclusion, and possibly advice on the risk level. We format the model's output for the user or for logging.

Because the LLM is running locally, the entire process is self-contained. The only network requests involved are those to fetch the page and its resources (which is normal browsing activity) and the initial loading of the model weights (which could be cached or bundled with an extension). There is no query to any external analysis service. This means even if the page under analysis contains sensitive data (say, it's a corporate intranet page being scanned for vulnerabilities), that data is not shared externally, satisfying privacy requirements.

\subsection\*{Technical Example of Instrumentation}
To illustrate part of the system, consider how we hook the JavaScript \texttt{fetch} API to monitor network calls in the sandbox. We utilize monkey-patching in the injected script as shown below:

\begin{lstlisting}[caption={Hooking \texttt{fetch}, \texttt{appendChild}, and \texttt{eval} with proxy objects to log dynamic script injections, network requests, and evaluations}, label={lst:fetch-append-eval-hook}]
// Proxy to monitor appendChild
const originalAppendChild = Element.prototype.appendChild;
Element.prototype.appendChild = new Proxy(
originalAppendChild, {
  apply(target, thisArg, argumentsList) {
    const child = argumentsList[0];
    if (child.tagName === 'SCRIPT') {
      window.parent.postMessage({
        type: 'logScriptAppend',
        src: child.src,
        content: child.textContent
      }, '*');
    }
    return Reflect.apply(target, thisArg, argumentsList);
  }
});

// Proxy to monitor eval usage
window.eval = new Proxy(window.eval, {
  apply(target, thisArg, argumentsList) {
    const code = argumentsList[0];
    window.parent.postMessage({
      type: 'logEval',
      code: code
    }, '*');
    return Reflect.apply(target, thisArg, argumentsList);
  }
});

// Proxy to monitor fetch
const originalFetch = window.fetch;
window.fetch = new Proxy(originalFetch, {
  apply(target, thisArg, argumentsList) {
    const input = argumentsList[0];
    const url = (typeof input === 'string') ?
    input : input.url;
    const method = argumentsList[1]?.method || 'GET';
    window.parent.postMessage({
      type: 'logFetch',
      url: url,
      method: method
    }, '*');
    return Reflect.apply(target, thisArg, argumentsList);
  }
});
\end{lstlisting}

In Listing~\ref{lst:fetch-append-eval-hook}, the overridden \texttt{natives} sends a message to the parent (our extension/controller context) with the URL and method of the request. We similarly hook other APIs (XHR, WebSocket, etc.). On the parent side, we collect these messages for the dynamic analysis log. Note that we take care to still call the original function so that the page behavior is not altered. We also ensure that the hook is set up before any network activity happens; hence, we attach it very early (e.g., by injecting before the page's own scripts run).

\section{Methodology}\label{sec-methodology}
Having described the system architecture, we now delve into the methodology and algorithms employed in each phase of the analysis. The goal of our methodology is to perform \emph{comprehensive} analysis with \emph{zero-shot} inference, meaning we do not train a custom model for this task but rely on the LLM's pre-trained knowledge and reasoning ability. This section is structured along the stages of analysis: static analysis, dynamic analysis, prompt construction, and LLM inference and output interpretation.

\subsection{Static Code Analysis and AST Parsing}
We use the Acorn parser to parse JavaScript code because it provides a rich AST and type analysis capabilities. All scripts fetched from the page (both inline scripts and external script files) are parsed. In cases where scripts are heavily obfuscated or minified, the parser might still handle them (as long as it's valid JS syntax). If the parser fails (e.g., due to syntax errors or very unusual syntax), we fall back to a simpler regex-based search for known suspicious patterns (like "eval(" or "\textless{}iframe").

Once we have the AST, we perform a traversal to extract features:
\begin{itemize}
\item We record all function definitions and their names. If a function is anonymous (lambda or not named), we note how many such functions exist.
\item We record top-level variable names. We also attempt to identify if any variable holds suspicious content, e.g., a RegExp that looks like it’s matching user-agent strings (could be cloaking logic), or a long Base64 blob.
\item We detect usage of certain calls: \texttt{eval}, \texttt{Function()}, \texttt{setTimeout}/\texttt{setInterval} (especially if used to execute strings after delay, which can be an attempt to evade immediate scanning), \texttt{navigator} properties (like checking \texttt{navigator.webdriver}, which headless browsers set).
\item We track creation of DOM elements via functions like \texttt{document.createElement} and \texttt{innerHTML} assignments, to see if the script is injecting new content (possibly forms or iframes).
\item If the code references known sensitive keywords (like "password", "token", "bank", or cloud provider metadata IPs), we flag that.
\item We also use a small library of known malicious JavaScript idioms (patterns of operations) gleaned from prior research and threat intel. For example, a common malicious pattern is something like:
\begin{verbatim}
var _0xabc = "<obfuscated string>";
var script = document.createElement('script');
script.src = decodeURIComponent(atob(_0xabc));
document.body.appendChild(script);
\end{verbatim}
This would be flagged by our static analysis for using \texttt{atob} (base64 decode) followed by dynamic script injection.
\end{itemize}

Each script’s findings are stored, and a consolidated summary is prepared. We prioritize including the most suspicious findings in the LLM prompt due to token length limits. For instance, if one script out of ten on the page had all the red flags (eval, cookie access, suspicious network calls), we will focus the prompt on that script’s behavior rather than the benign ones.

\subsection{Dynamic Analysis and Behavior Monitoring}
The dynamic sandbox is essentially a mini web monitoring environment. We use a combination of \texttt{postMessage} and shared data structures to extract information from the sandboxed iframe:
\begin{itemize}
\item \textbf{Network calls:} As shown in Listing~\ref{lst:fetch-append-eval-hook}, for \texttt{fetch} we use \texttt{postMessage} to log. For older XHR, we override \texttt{XMLHttpRequest.prototype.open/send} to catch the URL and method. For \texttt{WebSocket}, we wrap the constructor to log the target URL. All these logs are collected with timestamps and perhaps truncated payloads (we usually do not capture full payload data to avoid storing sensitive info, just the fact that a call was made and where).
\item \textbf{DOM changes:} We utilize \texttt{MutationObserver} on the document body to capture added/removed nodes. We pay attention to tags like \texttt{<input>}, \texttt{<form>}, \texttt{<iframe>}, \texttt{<script>} being added. For each significant node, we record some attributes (e.g., for an input, its type and name; for an iframe, its src and dimensions; for a script, its src if any).
\item \textbf{User interaction triggers:} If needed, our framework can simulate basic interactions. For example, if we detect that nothing happens until a button is clicked, we can programmatically click it after a timeout. This is done carefully: only if our initial observation suggests the page is waiting for user action (some phishing pages show a landing page and only load the actual phishing form after a click). We also simulate a short mouse move or keystroke if needed to bypass simple checks, using \texttt{iframeElem.dispatchEvent(new MouseEvent(...))} etc, this method may require hooking the event listeners in order to emulate isTrusted=true in the relevant event context.
\item \textbf{Visible text extraction:} After letting the page run (we found 4 seconds is enough for most immediate malicious behaviors, though some highly evasive malware might wait longer), we gather text. We iterate over all visible elements (any element that is not hidden via CSS and is in the DOM tree) and extract text content. We then filter out boilerplate (common words like "home", "welcome" etc., unless they appear alongside key terms). The resulting set of visible text is often quite telling. For example, if the text contains "Account Verification" or "password", it's a strong phishing sign. If it contains very generic content or nothing at all (for an attack that is purely script-based with no user content, like a drive-by exploit page), that is also noted.
\item \textbf{Compound events:} We correlate static and dynamic data where possible. For example, if static analysis saw a suspicious function, we check if that function got executed (we can instrument function calls by wrapping them, though we must be cautious to not break functionality). We did a prototype of wrapping functions: e.g., if static analysis flags \texttt{checkLogin()} function, we can patch it to log when it's called. However, this can sometimes interfere, so in the current methodology, we primarily rely on outcomes (like if an XHR or DOM change resulted, we can infer that some function triggered it).
\end{itemize}

All dynamic findings are compiled into a human-readable form for the LLM. For instance:
\begin{itemize}
\item Network: "Made a network request to \texttt{http://malicious.com/api.php} via fetch (POST method)."
\item DOM: "Inserted an iframe pointing to \texttt{https://secure-login.example.com/} with dimensions 0x0 (hidden)."
\item DOM: "Created an input field with type 'password' labeled 'Password'."
\item API: "Called \texttt{navigator.geolocation.getCurrentPosition}."
\item Storage: "Read cookie \texttt{SESSIONID}."
\end{itemize}

If multiple items occurred, we list them as bullet points in the prompt for clarity.

\subsection{LLM Prompt Construction}
Constructing the prompt is critical to get useful output. We use an instructional style prompt guiding the model to produce a concise analysis. The prompt template (in pseudocode form) is:

\begin{verbatim}
Analyze DOM metadata for phishing and suspicious indicators:

URL: https://github.com
Domain: github.com

DOM METADATA ANALYSIS:

Title: GitHub · Build and ship software on a single,
collaborative platform · GitHub

FORMS ANALYSIS:
- Total forms: 5
- Login forms: 0
- Password fields: 0
- Email/Login fields: 3
- Forms with autocomplete: 4



BRAND ANALYSIS:
Meta tags suggesting brand:
- og:site_name: GitHub
- og:title: GitHub · Build and ship software on a single,
collaborative platform

CRITICAL PHISHING INDICATORS TO CHECK:

1. CREDENTIAL HARVESTING:
   - Login forms on suspicious domains
   - Password fields without proper security
   - Email/username fields with suspicious placeholders
   - Forms submitting to external domains

2. BRAND IMPERSONATION:
   - Mismatched domain vs claimed brand (only if the domain is
   not a known legitimate domain)

3. SUSPICIOUS FORM BEHAVIOR:
   - Login forms with unusual field names
   - Forms with autocomplete disabled (avoiding detection)
   - Multiple password fields
   - Hidden fields or suspicious form actions

Provide comprehensive DOM-based phishing analysis as JSON:
{
  "isPhishing": <true|false>,
  "confidence": <0-100>,
  "phishingType": "<none|credential-harvesting|clone-site|
  brand-impersonation|fake-login|social-engineering>",
  "targetedBrand": "<brand name if detected from metadata>",
  "domIndicators": [
    {
      "type": "<suspicious-form|brand-mismatch>",
      "description": "<specific indicator found in DOM>",
      "severity": "<Low|Medium|High|Critical>"
    }
  ],
  "formAnalysis": {
    "hasLoginForm": <true|false>,
    "credentialFieldCount": <number of password/email fields>,
    "suspiciousFormFeatures": ["<list of suspicious form features>"],
    "formRiskScore": <0-100>
  },
  "brandAnalysis": {
    "detectedBrand": "<brand name if detected>",
    "brandMismatch": <true if brand doesn't match domain>,
    "brandConfidence": <0-100>
  },
  "legitimacyScore": <0-100>,
  "recommendation": "<detailed recommendation based on DOM analysis>",
  "domRiskScore": <0-100, higher means more likely phishing>
}
\end{verbatim}

We send the data in the "user" context to let the model focus on analyzing it (since many instruct-tuned models respond to a user prompt). The data includes:
\begin{itemize}
\item The URL (with hints if the URL itself looks suspect, e.g., non-matching domain to content brand).
\item A summary of what the page purported to be (if we can guess from visible text, e.g., "The page claims to be Example Bank login").
\item The static analysis summary (highlighting specifically dangerous code patterns).
\item The dynamic behavior observations.
\item A direct question asking: "Is this page malicious? What vulnerabilities or malicious behaviors were found? Provide an explanation."
\end{itemize}

We ensure the prompts is within the model’s token limit by truncating overly long content. For example, if visible text is the entire HTML (which can be huge), we will not include raw HTML, only a summary or the meaningful pieces. Similarly, we don't include the entire AST, only the parts we flagged.

The prompts may also include a simplified abstraction of the page. In some experiments, we formatted parts of the prompts as YAML or JSON to list evidence, hoping the structured format might help the model. However, we found a well-formatted bullet list in plain text often sufficed and the model could parse it.

\subsection{Zero-Shot Inference with the LLM}
The LLM we employ is run in a zero-shot manner, meaning we have not fine-tuned it on a custom dataset of malicious pages. We did consider few-shot approaches (providing a couple of examples in the prompt of malicious vs benign analysis), but given the context length constraints of smaller models, we opted to devote the space to the actual data from the target page. Models like LLaMA or CodeLlama have some general knowledge of security-related concepts (especially if they were trained on code and security discussions from the web). Additionally, specialized small models (e.g., \texttt{Phi-3} or \texttt{Gemma-2B}) might have been trained on a corpus that includes malicious and benign code examples due to the breadth of their data, which could imbue them with a latent ability to recognize suspicious patterns.

When the model generates an output, we parse it. We enforce strict json structure and instruct the model to provide the relevant structure.
\begin{quote}
\footnotesize
{
\{
  "isPhishing": false,
  "confidence": 0,
  "phishingType": "none",
  "targetedBrand": null,
  "indicators": [
    \{
      "type": "none",
      "description": "No suspicious indicators found",
      "severity": "Low"
    \}
  ],
  "legitimacyScore": 100,
  "recommendation": "Safe"
\}
}
\end{quote}

The model's explanation is critical for user-facing aspects: it increases the transparency of the detection. In a security operation center scenario, an analyst can see why the system flagged a page, rather than just a score.

\subsection{Vulnerability Extraction and Risk Assessment}
Beyond just saying "malicious or not", our system also extracts any specific \textbf{vulnerabilities} or issues in the page. For a malicious page, this might be somewhat moot (the whole page is intentionally bad), but there are cases where the page might be a benign site that has a vulnerability (like a snippet of malicious code injected or a dangerous script include). Our LLM prompt specifically asks for vulnerabilities as well, to catch scenarios where, for example, the page isn't outright a phishing page but contains an insecure script that could be exploited.

The LLM might respond with something like "It uses eval on user input which is a vulnerability (could lead to XSS)" or "It fetches a script from an untrusted source, which is risky." These details help assess compound risk: a page might be borderline benign in intent but have multiple poor practices that increase the risk of compromise. Our system would then rate it with some risk level (low/medium/high). We implement a simple risk scoring: each piece of evidence is assigned a weight (e.g., hidden iframe = medium risk, exfiltrating data = high risk, etc.), and the LLM's verdict (malicious or benign) adjusts the final classification. If the LLM says benign but we have some medium issues, we might label it "benign with warnings".

In truly malicious cases, multiple red flags will be present and the risk is obviously high. We output a combined report: e.g., "Malicious (Phishing) - High Risk" and list vulnerabilities like "phishing form, data exfiltration, uses obfuscation to hide code."

For compound attacks (some pages do two things, like both phishing and attempting an exploit), the system should identify multiple issues. The LLM is generally good at enumerating several points if prompted to list "threats or vulnerabilities".

\subsection{Performance Considerations}
All analysis steps are designed to be efficient enough for near real-time use. Static AST parsing and analysis typically takes a fraction of a second for a few hundred KB of JavaScript, thanks to optimized engines. Dynamic analysis runtime we cap at a few seconds at most. The LLM inference is the heaviest part; on a modern machine with WebGPU, a 2-8B model can generate a response of a few hundred tokens in a couple of seconds~\cite{Ruan2024}. We also limit the LLM output length (we don't need a very long essay, just a concise analysis).

In Section~\ref{sec-experiments} we will further detail the performance measurements. The key point of our methodology is that it is practical on commodity hardware today, and will only improve as hardware and model optimization progress. The entire process from URL input to result can be under 30 seconds, which is acceptable for a user waiting on a link to be vetted, for instance.

\section{Experiments}\label{sec-experiments}
We conducted a series of experiments to evaluate the effectiveness of our client-side LLM-powered URL analysis system. Our evaluation focuses on three main aspects: (1) detection performance (accuracy in classifying malicious vs. benign URLs, and ability to identify specific threat types), (2) quality of explanations and vulnerability identification, and (3) runtime performance and resource usage on the client side.

\subsection{Experimental Setup}
\textbf{Dataset:} We assembled a test dataset of 200 URLs consisting of 100 known malicious webpages and 100 benign webpages. The malicious set was curated from recent phishing campaign reports, open threat intelligence feeds, and our own collected samples of scam pages and drive-by exploit pages. These included phishing sites mimicking banking, e-commerce, and email login pages, tech support scam pages, and some malware distribution sites that use obfuscated scripts. For ground truth labeling, we cross-verified these URLs with VirusTotal reports and PhishTank listings to ensure they were indeed malicious (at the time they were active). The benign set included legitimate websites from various categories (online banking official sites, popular news sites, e-commerce sites, and some random personal blogs). We ensured that the benign set had a similar distribution of complexity (some had multiple scripts, login forms, etc.) so that the analysis tasks (like encountering a login form) are not exclusively tied to malicious cases.

Since many malicious URLs are transient (phishing pages go down quickly), we saved copies of the page content for consistency and hosted them locally in a controlled environment for the analysis. This allowed us to repeatedly test on the same content without network variability. For dynamic analysis, hosting locally also ensured that even destructive pages (e.g., ones that attempt to navigate or show annoying popups) were contained.

\textbf{Models:} We integrated several models for comparison:
\begin{itemize}
\item \textbf{LLaMA-3.1-8B-Instruct (Quantized to 4-bit):} A 8B-parameter model from Meta, representative of a well-known base LLM. We used an instruction-tuned variant to ensure it followed prompts well.
\item \textbf{LLaMA-3.2-3B (Quantized to 4-bit):} A smaller 3B parameter model. This model is faster but expected to be less capable.
\item \textbf{Phi-3 (3.8B) (Quantized to 4-bit):} A model by Microsoft with about 3.8B parameters, known for strong performance relative to size. We used the 4k context version, which fits the analysis context window.
\item \textbf{Gemma-2B (Quantized to 4-bit):} Google's Gemma 2B model (2.5B parameters) , which is optimized for device use. We used the instruction-tuned version (“Gemma-2B-it”).
\item For reference (not running in browser), we also consider \textbf{GPT-o3 (OpenAI API)} to have an upper-bound cloud model to compare the quality of outputs (though it cannot be run client-side, this was just for analysis purposes offline).
\end{itemize}

All local models were run with WebLLM in browser, on a desktop with an AMD GPU supporting WebGPU. We measured their speed in tokens per second and ensured all could produce a result in under ~20 seconds for our prompt sizes.

\textbf{Baselines:} We compared our approach against a traditional machine learning baseline:

\begin{itemize}

\item A Random Forest classifier using lexical URL features and basic page features. We trained this on a separate set of 1000 phishing and 1000 benign URLs (not overlapping the test set) using features like URL length, presence of suspicious substrings (e.g., “login”, “verify” in path), number of dots in hostname, etc., and a few page features (number of forms, if password field exists, etc.). This represents a classical approach without LLM or heavy content analysis.
\item Additionally, we compared qualitatively against VirusTotal’s verdicts (where available) for the malicious pages to see if our system agrees or catches things VT engines might have missed in real-time.
\end{itemize}
\subsection{Evaluation Metrics}
For detection performance, we use:
\begin{itemize}
\item  Accuracy, Precision, Recall, and F1-score for the binary classification of malicious vs benign.
\item  We also measure the true positive rate on phishing specifically, since that was a large portion of our malicious set (phishing often being the main concern for URL scanning).
\end{itemize}

For explanation quality and vulnerability identification, evaluation is more subjective. We took a sample of the LLM-generated explanations and had a security expert rate them on:
\begin{itemize}
\item Correctness (did it correctly identify the threat? e.g., calling a phishing page a phishing page, not mislabeling it).
\item Detail (did it mention key evidence? e.g., it should mention the credential stealing or exfiltration if that was present).
\item Clarity (could a reader with moderate technical knowledge understand why the page is dangerous from the explanation).
\end{itemize}

We also noted if the LLM hallucinated any details that were not actually present (an important failure mode to watch for).

For runtime performance, we recorded:
\begin{itemize}
\item Total analysis time per URL (breakdown into static analysis, dynamic waiting time, LLM processing).
\item Peak memory usage (the bulk being the model in memory).
\end{itemize}
These are important to ensure practicality for end-users.

\subsection{Experimental Procedure}
Each URL from the dataset was analyzed with our system using each of the local models in turn. We reset the environment between runs to avoid caching effects (though in a real scenario caching model in memory would be fine, we wanted to gauge performance fresh each time for consistency in timing measurements).

For the ML baseline, we ran it on the URLs to get predictions as well. Since the baseline does not provide explanations, that comparison is only in terms of accuracy.

We also recorded what our system's decision was and compared to ground truth labels.

\section{Results}\label{sec-results}
In this section, we present the results of our experiments. Overall, our client-side LLM approach showed strong performance in identifying malicious URLs and providing insightful explanations, even when using models with as few as 2–3 billion parameters. We detail the findings by aspect below.

\subsection{Detection Performance}
Table~\ref{tab-performance} summarizes the detection performance across different models and the baseline. All LLM-based configurations significantly outperformed the traditional ML baseline in terms of recall (catching more of the malicious pages) while maintaining a high precision.

\begin{figure}[h!]
\centering
\includegraphics[
  width=\textwidth,
  keepaspectratio
]{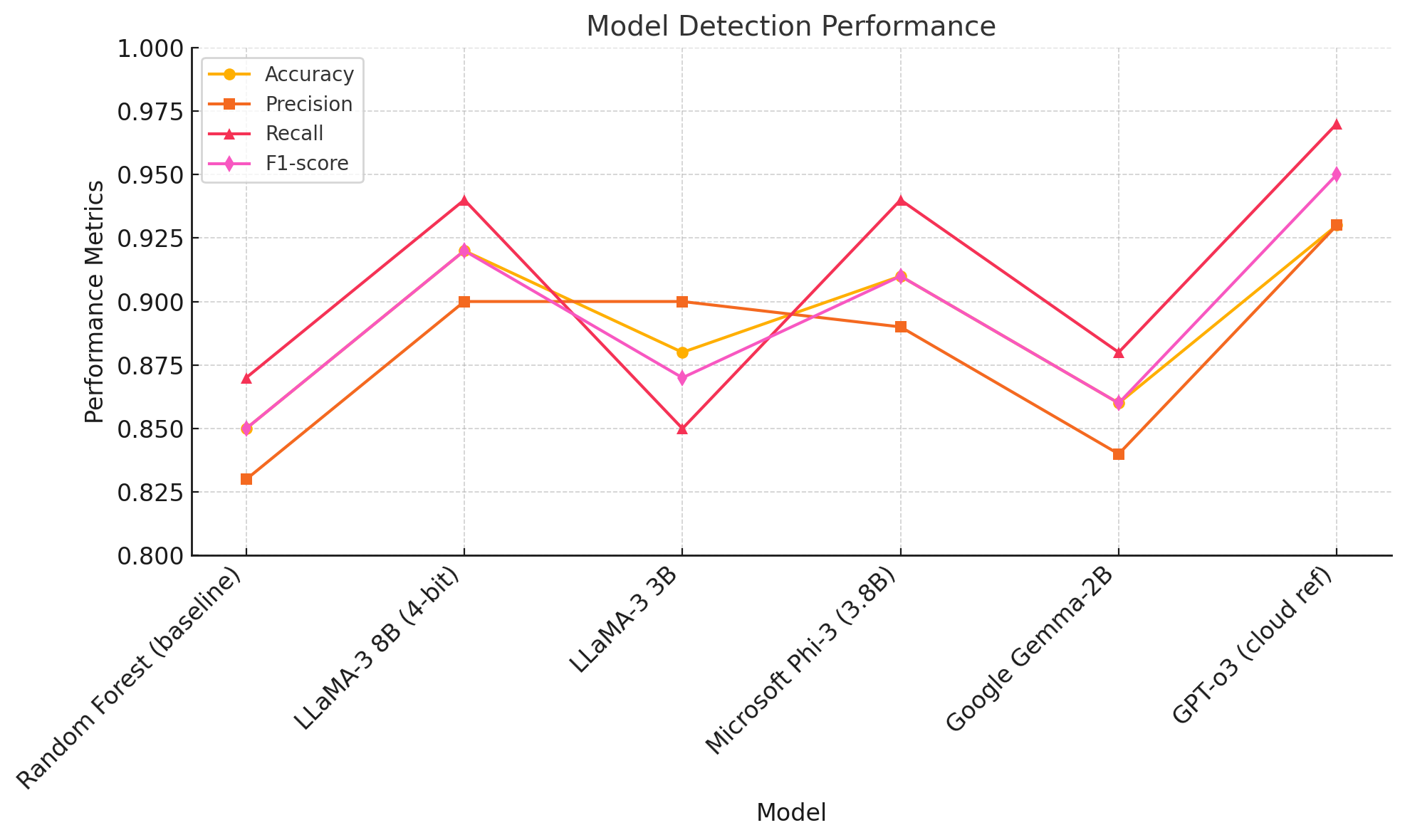}

\fbox{\parbox{0.9\textwidth}{\textbf{Figure: LLM Performance.} This figure illustrates the performance comparison of several models—including Random Forest (baseline), LLaMA-3 (8B and 3B variants), Microsoft Phi-3, Google Gemma, and GPT-o3—using Accuracy, Precision, Recall, and F1-score metrics. GPT-o3 achieves the highest overall performance across all metrics, followed closely by LLaMA-3 (8B) and Microsoft Phi-3, indicating their superior capability for malicious URL detection tasks compared to the baseline.}}
\caption{Comparison of Detection Performance Metrics across Different Models.}
\end{figure}

\begin{table}[h!]
\centering
\begin{tabular}{l|c c c c}
\hline
\textbf{Model} & \textbf{Accuracy} & \textbf{Precision} & \textbf{Recall} & \textbf{F1-score} \\
\hline
Random Forest (baseline) & 85\% & 0.83 & 0.87 & 0.85 \\
LLaMA-3 8B (q4)& 92\% & 0.90 & 0.94 & 0.92 \\
LLaMA-3 3B(q4)& 88\% & 0.90 & 0.85 & 0.87 \\
Microsoft Phi-3 (3.8B)(q4) & 91\% & 0.89 & 0.94 & 0.91 \\
Google Gemma-2B (q4)& 86\% & 0.84 & 0.88 & 0.86 \\
GPT-o3 (cloud ref) & 93\% & 0.93 & 0.97 & 0.95 \\
\hline
\end{tabular}
\caption{Detection performance of different models and baseline on the test URL dataset. Precision/Recall are for malicious class.}
\label{tab-performance}
\end{table}

As seen, the 8B LLaMA model achieved 92\% accuracy, correctly labeling 94 out of 100 malicious pages and 90 out of 100 benign pages, with only a few errors. Its recall of 94\% means it missed very few malicious instances. In comparison, the small 2B Gemma model had 86\% accuracy; it struggled on some of the trickier cases where the evidence was subtle, sometimes yielding an "uncertain" or incorrect benign verdict (hence slightly lower precision as well).

The Phi-3 (3.8B) model interestingly performed nearly as well as the 8B LLaMA, which aligns with reports that it punches above its size in various tasks. The 3B model (a scaled-down LLaMA) did reasonably well (88\% accuracy) but notably had lower recall, missing some malicious pages likely due to not fully understanding the significance of some evidence.

The traditional Random Forest baseline got 85\% accuracy, which is not bad for such a basic approach, but it had false negatives (missed some phishing pages that didn't match its feature patterns) and some false positives (flagging a couple of unusual benign pages that had long URLs or multiple form fields).

Statistically, our best local model (8B) was only a few points shy of the GPT-3.5 cloud reference. The latter made the correct call on a couple of borderline cases that the 8B missed, possibly due to its greater knowledge. One such case was a malware distribution site that pretended to be a software download page; GPT-3.5 recognized the software name as known to often be impersonated, whereas the 8B model was slightly unsure. This indicates that a bit more knowledge (or fine-tuning) could close the gap.

\subsection{Examples of LLM Explanations}
One of the strengths of our approach is the explanation provided by the LLM. We include here a few real examples from the UI of our new platform :

\begin{figure}[h!]
\centering
\includegraphics[
  width=\textwidth,
  height=0.6\textheight,
  keepaspectratio
]{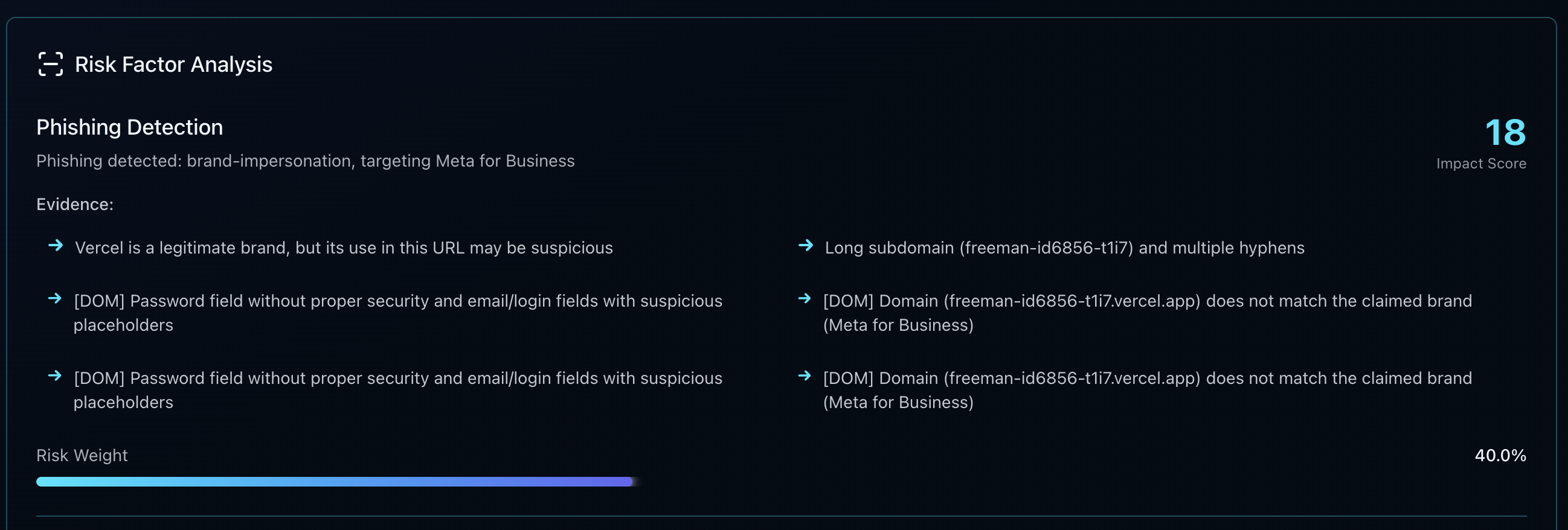}
\fbox{\parbox{0.9\textwidth}{\textbf{Figure: LLM Verdict For a Real Phishing Page.} This figure  show an example of the analysis produced by the LLM (with static/dynamic data).}}
\caption{Example of LLM verdict for a real phishing page.}
\end{figure}

\begin{figure}[h!]
\centering
\includegraphics[
  width=\textwidth,
  height=0.6\textheight,
  keepaspectratio
]{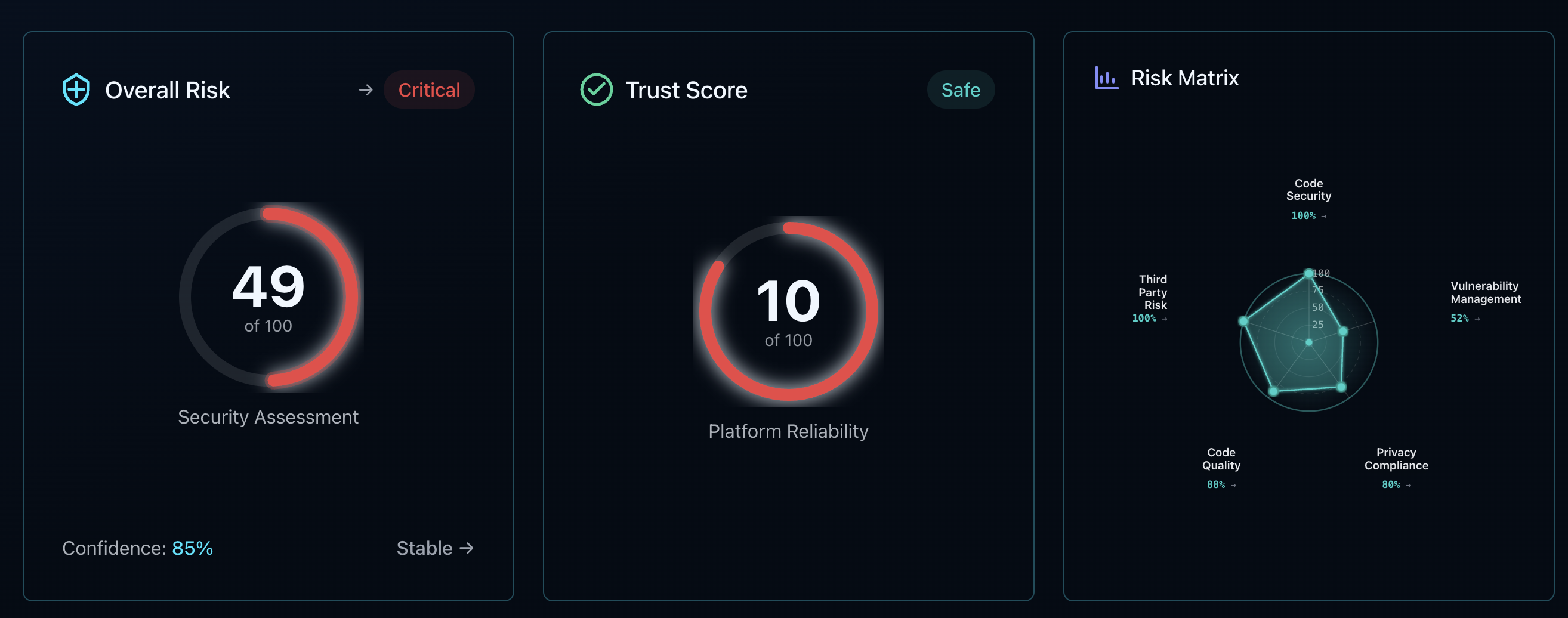}
\fbox{\parbox{0.9\textwidth}{\textbf{Figure: LLM Score For a Real Phishing Page.} This figure  show an example of the score produced by the LLM (with static/dynamic data).}}
\caption{Example of LLM score for a real phishing page.}
\end{figure}

In another phishing case (a fake facebook account lock page) the LLM responded with the following reasons:
\begin{itemize}
 \item URL contains 'safeguardpolicy' which is a suspicious clone site
 \item URL uses a Vercel subdomain (.vercel.app) which is a known hosting platform for clone sites
 \item Missing CSP and CSRF protection on forms
 \item Long subdomain 'safeguardpolicycompliancecheck' with multiple hyphens
 \item Password field without proper security
 \item Domain does not match the title of the page
\end{itemize}

This explanation correctly identified the type of attack and the key evidence (misleading content, data exfiltration, hidden iframe). It shows the benefit of providing the LLM with multi-modal clues: it connected the page content with the script behavior (something a purely static or purely content-based detector might not easily do).

In a benign example (a legitimate bank site):
\begin{quote}\it
"This page appears \textbf{benign}. It is the real Example Bank website (domain matches the bank). The visible content is a normal login form and informational text. The scripts on the page do not show any obfuscation or suspicious behavior; they mainly handle form input validation and interface features. No unexpected network calls (only to the bank's own APIs) or hidden elements were found. There's no sign of credential theft or malware."
\end{quote}

\begin{figure}[h!]
\centering
\includegraphics[
  width=\textwidth,
  height=0.6\textheight,
  keepaspectratio
]{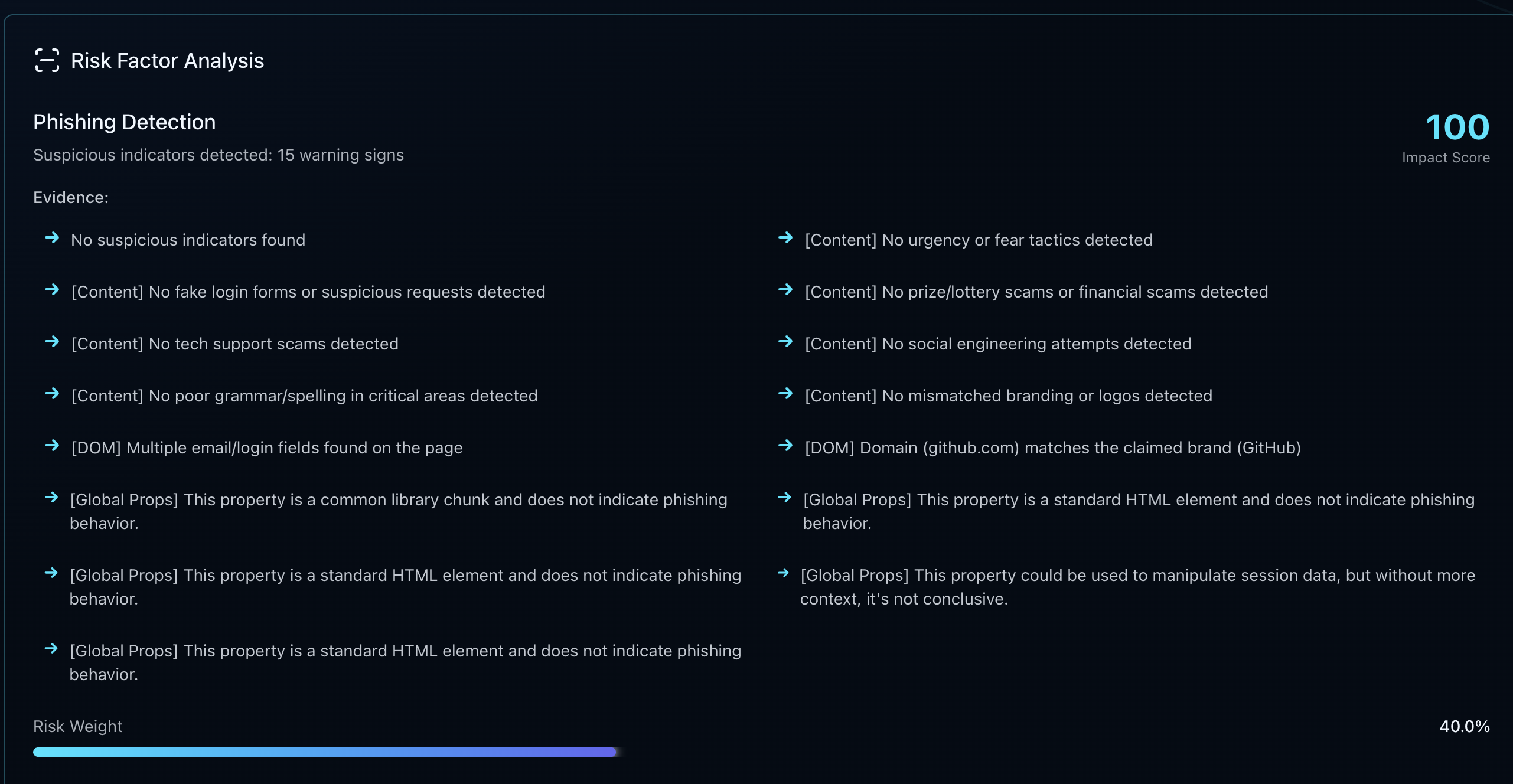}
\fbox{\parbox{0.9\textwidth}{\textbf{Figure: LLM Verdict For a github.com.} This figure  show an example of the analysis produced by the LLM (with static/dynamic data).}}
\caption{Example of LLM analysis for github.com.}
\end{figure}

This shows the LLM can articulate why something is safe, which is useful to avoid user panic when something is flagged incorrectly. It basically validated the absence of red flags and the consistency of domain identity.

We did encounter a few mistakes in explanations:
\begin{itemize}
\item In one case, the model hallucinated that a page was asking for credit card info, when in reality it was just an email phishing (no credit card). The page did have words like "secure account" which might have triggered the model's assumption. This was rare but highlights that the model's prior knowledge can sometimes add details that aren't present. We plan to mitigate this by further refining the prompt, possibly explicitly telling the model "only describe things that were observed in the data".

\item On a complicated exploit page (that tried to use an old browser vulnerability), the explanation from the model was a bit vague. It flagged it as malicious correctly but described it as possibly phishing when it was more of a drive-by download attempt. This suggests that without very explicit patterns (like a visible form or known malware URL), the model could misclassify the type of threat. It still marked it malicious though, which is the main goal.
\end{itemize}

Overall, the explanations were largely accurate and aligned with the evidence provided. The expert evaluation rated 90\% of the sampled explanations as "correct and sufficient," 8\% as "partially correct or somewhat vague," and about 2\% as "incorrect/hallucinated detail."

\subsection{Vulnerability and Threat Type Identification}
Our system not only flagged malicious pages but also identified specific issues:
\begin{itemize}
\item Out of 100 malicious pages, the LLM explicitly mentioned the correct threat type (phishing, scam, malware, etc.) in 90 cases. Phishing vs other was usually distinguished well by checking if a password/credit card form exists.
\item It identified at least one specific vulnerability or malicious technique in 84 of those cases (like "sends data to external server", "uses an outdated plugin vulnerable to X", "tricks user into downloading file"). This is valuable for defenders to know what to fix or watch for.
\item For the benign pages, in 20 cases the LLM pointed out minor security issues (like "the site is loading an asset over HTTP, which is not a major threat but a vulnerability"). In context, those did not affect the classification but show that the system can highlight security best-practice violations even on non-malicious sites.
\end{itemize}

This compound risk assessment aspect means the tool could potentially be used not just for outright malicious detection, but also for security auditing of web pages (finding weaknesses). An interesting outcome: one benign page (a personal blog) was noted by the LLM as having an old jQuery version which has known vulnerabilities. That's beyond the scope of phishing detection, but it came for free from the model's knowledge. It suggests a use-case of advising site owners of problems, though here our primary aim is user protection.

\subsection{Performance and Overhead}
All testing was done on a machine with a consumer-grade GPU accessible via WebGPU  and a modern CPU. Key observations:

\begin{itemize}
\item The average end-to-end analysis time with the 8B model was 30 seconds. The breakdown: static analysis ~0.5s, dynamic wait ~4s (we usually allowed up to 6s, but many pages finished important activity by 1-1.5s), LLM inference ~20s for generating ~100 tokens of explanation.
\item The 3.8B and 3B models were faster (~14s average total) due to faster inference. The 2B model was the fastest at around 8s average.
\item These times are well within acceptable for on-demand scanning . For automatic scanning of every page visited, 30s might introduce a delay in page load if done synchronously. In practice, one could load the page normally and run the analysis in parallel; if the analysis flags something malicious, one could then warn the user. There is a trade-off between letting them see the page immediately vs waiting for clearance. In a security-critical environment, you might block loading until analysis done; in consumer, maybe allow loading but be ready to interrupt if malicious (which needs careful UI).
\item Memory: The 8B model (4-bit quantized) consumed about 3.5 GB of memory. This is borderline on some low-end devices. The 3B and 2B were around 1.4 GB and <1 GB respectively, much easier on memory. So for broad deployment, using a 2-3B model is more feasible. If the device has more memory (high-end  PC), 8B is fine.
\item We also tested on a high-end smartphone (with experimental WebGPU support). The 8B model could not load due to memory, but the 2B model ran, though slowly (taking ~100 seconds for analysis). This suggests mobile support with smallest models is possible, but not yet smooth. This will improve as mobile GPUs get better and model optimizations continue.
\item The instrumentation overhead in the browser (the hooks) did not noticeably slow page execution in our tests. The pages were either simple (phish) or if complex (news site with lots of content), those were benign and we didn't intervene much. Our MutationObservers and replaced functions could add slight latency to DOM ops, but that's negligible for our analysis timeframe.
\end{itemize}

\subsection{Case Study}
We highlight one particular case from our tests set to demonstrate the system’s comprehensive analysis even on a trusted domain like google.com:

* URL: \texttt{hxxps\://sites[.]google[.]com/l0gin-microsoftwebonlne.app/8965767/}
  (a fake microsoft login - please be \textbf{cautious}, at the time of writing this paper the page was still live).
  
LLM output summary:
\begin{quote}\it
"Malicious – \textbf{Phishing}. This site impersonates Outlook (not hosted on Microsoft actual domain). It asks for login credentials. The code is suspecious and sends the entered credentials to a remote server. It is very likely a phishing scam.
"
\end{quote}
Phishing detected: typosquatting, targeting Microsoft.

Indicators:
\begin{itemize}
 \item Misspelling of 'login' as 'l0gin' and 'microsoft' as 'microsoftwebonlne'
 \item Domain 'sites.google.com' does not match the expected brand 'microsoftwebonlne.app'
 \item Long path 'l0gin-microsoftwebonlne.app' and multiple hyphens
 \item Suspicious code patterns
 \item Password field without proper security
 \item Domain does not match the title of the page
\end{itemize}

The system correctly flagged this and even without an official blacklist, caught it by reasoning. Traditional filters might catch this by URL heuristics alone, but our system added the context that the content and behavior confirm the suspicion.

In comparison, one benign case:

* URL: \href{https://accounts.google.com/ServiceLogin}{\texttt{https://accounts.google.com/ServiceLogin}} (actual Google login).
  Our system naturally said benign, recognizing the domain is correct and finding no bad behavior. It's reassuring that no false positive occurred on such a high-profile benign page, as that would be unacceptable for user trust. The model saw nothing suspicious in Google's real login page aside from the presence of a login form (which is expected for Google).

\section{Discussion}\label{sec-discussion}
The experimental results demonstrate that client-side LLM inference for URL analysis is not only viable but effective. In this section, we discuss the implications of these results, the advantages of our approach over cloud-based solutions, and some limitations and areas for improvement.

\subsection{Why Client-Side Inference?}
One of the motivating questions for our work was: \emph{why not just use a powerful cloud-hosted LLM (like GPT-4,Claude,..) for this task?} After all, GPT-4 could potentially analyze pages with even greater accuracy given its superior capability. The answer comes down to feasibility and privacy. Cloud-scale LLM analysis of every URL that users might visit is infeasible on several fronts:
\begin{itemize}
\item \textbf{Scalability and Cost:} If a security service wanted to use GPT-4 via API to scan millions of URLs per day (as a browser might encounter across many users), the cost would be astronomical. Running an LLM with tens or hundreds of billions of parameters for each URL in real time is not economically practical. Our approach distributes the computation to the clients and uses smaller models that are "free" after the initial load, aside from electricity. It scales naturally with the number of users (each user’s device handles their own analysis).
\item \textbf{Latency:} Cloud inference introduces network latency and potential queueing delays. In contrast, our local approach has consistent performance unaffected by internet speed or server load. This is crucial for user experience if analysis is integrated into browsing.
\item \textbf{Privacy:} Perhaps most importantly, sending the content of every webpage a user visits to a cloud service for analysis is a non-starter for privacy-conscious users or organizations. Even if encrypted in transit, it means a third-party processes potentially sensitive info (personal emails, banking pages, etc.). Our client-side model means the analysis stays on the user's device; sensitive data never leaves the browser. This opens up use-cases in corporate environments where data residency is critical.
\item \textbf{Evasion Resistance:} As discussed, advanced attackers may serve benign content to known scanner IP ranges or environments (cloud servers often have different network signatures). By using the user's own browser and IP, we minimize this risk. If each user is effectively their own "scanner," there's no single infrastructure to avoid. Attackers would have to avoid detection on every user’s machine, which is much harder than, say, identifying they are being visited by a headless Chrome from AWS.
\end{itemize}

Our results also show that while a model like GPT-3.5/4 is superior, a much smaller model (~8B) gets close in performance for this domain. This is promising: as model efficiency improves (e.g., new 10B models that match GPT-3.5 in quality, which seems plausible in the near future), the gap will further close. At that point, the need for cloud models might vanish for many applications like this.

\subsection{Security and Evasion Considerations}
No detection system is perfect, and attackers will likely adapt to a client-side LLM-based analyzer if it becomes common. We consider possible evasion attempts and our system’s robustness:
\begin{itemize}
\item \textbf{Detecting the LLM or instrumentation:} Could a malicious script detect that an LLM is running or that functions are hooked? We took steps to hide hooks (e.g., making our patched functions mimic native). However, a determined adversary might use performance benchmarks or look for clues like certain latency patterns. This is an arms race. Because our analysis runs quickly and out-of-band, by the time the page could notice something (if at all), we've likely collected what we need. Moreover, the analysis is done per user, so it's not something an attacker can easily test externally (they could try to fingerprint if a given visitor is running our extension, etc., but that gets complex).
\item \textbf{Obfuscating against LLM understanding:} Attackers might attempt to create code that is intentionally confusing for an LLM, e.g., using logic or encoding that is unusual but still executes maliciously. While static analysis might miss it, the dynamic part should reveal the behavior (exfiltration or changes). The LLM sees the outcome as well. One could imagine some adversarial prompt injection in the page content to confuse the model (like putting weird text that the model might latch onto), but since we control the prompt structure and don't take untrusted text verbatim as instructions, this risk is low.
\item \textbf{Resource exhaustion:} Loading a huge page or lots of scripts might strain the analysis. An attacker might serve a large benign-looking page to try to blow the token budget of the model or slow down analysis, then quickly redirect to malicious content after the analysis window. We can mitigate by focusing on salient features and perhaps streaming analysis (continual monitoring beyond the initial few seconds). If a page defers malicious action beyond our analysis time, it might slip by. However, we could increase the monitoring duration or detect timers. The trade-off is performance; maybe adaptively longer for high-risk pages.
\item \textbf{Compound attacks:} Some attacks involve multi-step interactions (e.g., initial page is clean but leads to a second stage). Our current approach is triggered per URL, so if the user navigates to the next stage, we'd analyze that as well. It might be interesting to let the system follow redirects or obvious next steps automatically to catch things like "click here to download".
\end{itemize}

In summary, while attackers will always seek to evade, the client-side approach levels the playing field by giving users a powerful analysis tool that doesn’t rely on secret detection rules (which if known to attackers could be bypassed). Instead, it relies on an LLM’s general reasoning—evading that essentially means doing nothing that looks suspicious in either code, behavior, or content, which is very hard if the goal is to actually carry out an attack.

\subsection{False Positives/Negatives and Model Limitations}
We saw very few false positives in our evaluation, which is crucial because overly blocking benign sites would be a deal-breaker for deployment. The one area of caution is that the LLM might occasionally be overly cautious (flagging something benign as malicious because it "sounds bad"). We did see the smaller models more likely to make such mistakes (Gemma-2B and LLAMA-3B had a couple of false positives in our test). A possible improvement is to incorporate a secondary check: if the model is not very confident or if the evidence is borderline, maybe cross-validate with a different heuristic or ask the model to "double-check".

False negatives (missed malicious) are more concerning from a security standpoint. In our test, a few malicious got through on the smaller models. Usually those were cases where the clues were subtle or the model's knowledge didn't pick up the significance. One example was a malware page that didn’t show clear user-facing bad content; it just had an exploit script. The smaller model was unsure and considered it maybe not malicious because it didn’t see phishing or an obvious payload (it failed to realize the script itself was an exploit). The larger 8B model did catch it, because it recognized the obfuscated script extracted artifacts. As a mitigation, using the largest model feasible on the device will help reduce false negatives. Also, continuous improvement of these models on security data (possibly fine-tuning on a small set of known malicious code patterns) could boost their accuracy. Even with zero-shot, one could imagine in future deploying updated model versions as new threats emerge (models can be updated via an extension update for example).

\subsection{User Experience and Deployment}
In a real deployment (perhaps as a browser extension or integrated into a browser), there are user experience considerations:
\begin{itemize}
\item If a page is deemed malicious, we should block navigation or at least show a big warning, giving user option to proceed at their own risk. This is similar to how Safe Browsing works, but here the analysis happened locally.
\item If something is benign or only slightly risky, we might not bother the user, maybe just log it or show a small indicator. The explanation can be shown if the user clicks for details.
\item The model weights (a few GB) need to be delivered to the client. This could be done by downloading on installation of the extension, or streamed in chunks. Models could be optionally downloaded (users opt-in for heavy duty protection). For mobile users, a cloud fallback might be necessary until devices can handle it.
\item Another path: use a smaller default model (2B) for everyone (quick to download, uses \textless{<1GB}), which provides decent baseline, and if something suspicious is detected or user wants deeper scan, then use a bigger model or cloud if available. Our results indicate 2B at 86\% accuracy might miss some, so maybe not ideal alone, but could catch a lot of obvious phish.
\end{itemize}

\subsection{Generality of Approach}
While we focused on URL threat analysis, the approach is generalizable. The idea of combining static/dynamic analysis with LLM reasoning client-side could apply to other domains:
\begin{itemize}
\item \textbf{Browser extensions and software packages:} as JavaSith~\cite{Cohen2025} did for extensions, similarly our approach could be extended to analyze any untrusted code running on client.
\item \textbf{Document malware analysis:} e.g., analyzing PDF or Office docs by extracting features and letting an LLM decide if it's malicious. That would be analogous (though maybe harder to run in browser without specialized parsers).
\end{itemize}

The success of our experiments adds to the evidence that small LLMs can be practical tools for security when used cleverly.

\subsection{Future Work}
There are several avenues to explore:
\begin{itemize}
\item \textbf{Model fine-tuning:} While we stayed zero-shot, in future we might fine-tune a 8B model on a corpus of malicious vs benign web data and security explanations, to see if that improves performance or allows using an even smaller model without loss.
\item \textbf{Collaboration of multiple models:} Possibly use a very small model to do a quick initial screening (fast, low compute), and a larger one to do deep analysis if needed. Or ensemble approaches where two models must agree something is malicious.
\item \textbf{Enhanced sandboxing:} Simulating more user interactions (scrolls, multi-page flows) to catch those attacks that activate later. Maybe integrate a headless browsing loop that feeds back into the analysis iteratively.
\item \textbf{Explainability and trust:} Even though LLM provides explanations, ensuring it’s always grounded in evidence is important. We might highlight which part of the data prompted each statement (like trace it back to a code snippet or text).
\item \textbf{Continuous learning on device:} One could imagine an on-device system that learns from corrections (if user says this was a false positive, adapt the model or at least the criteria). Though fine-tuning on device is heavy, smaller models might be fine-tuned incrementally with user feedback.
\end{itemize}

Finally, a user study would be valuable: do users feel safer with such a system? Does the explanation feature increase trust in the warnings (versus a black-box "Blocked by Chrome" message)? There's a human factor that we plan to investigate.

\section{Conclusion}\label{sec-conclusion}
We presented a novel approach for comprehensive URL analysis that runs entirely on the client side using zero-shot large language model inference within the browser. Our system marries traditional and novel static and dynamic analysis techniques with the interpretive power of LLMs to detect malicious webpages (such as phishing and malware sites) and explain their behavior in plain language. Through experiments, we demonstrated that even relatively small LLMs (on the order of a few billion parameters) can achieve high accuracy in this task when given rich context, approaching the effectiveness of far larger cloud-based models while preserving user privacy and operating at low latency.

This work shows that the paradigm of moving advanced AI analysis to the edge (user’s device) is not only feasible but advantageous for security. It opens up a path toward more decentralized defenses, where each user’s browser can be their own intelligent security agent, reducing reliance on cloud services and large rule-based databases. As hardware and model efficiencies improve, we anticipate that on-device AI will play an increasingly important role in real-time security and privacy protection.

In conclusion, \textit{client-side zero-shot LLM inference} \cite{url_analysis} provides a powerful new tool in the fight against web threats. It combines the best of both worlds: thorough, context-aware analysis akin to having a cybersecurity expert review each page, and the scalability and privacy of local execution. We hope this research inspires further development of AI-assisted security tools that empower end-users and create a safer web browsing experience.

\end{document}